\documentclass[pre,aps,twocolumn,showpacs]{revtex4}
\usepackage{amsmath,amssymb,graphicx}
\usepackage{epsfig}
\usepackage{textcomp}
\usepackage{color}

\begin{document}
\title{Velocity fluctuations of noisy reaction fronts propagating into a metastable state: testing theory in stochastic simulations}
\author{Evgeniy Khain$^1$ and Baruch Meerson$^2$ }
\affiliation{$^1$Department of Physics, Oakland University, Rochester, MI 48309, USA}
\affiliation{$^2$Racah Institute of Physics, Hebrew University of
Jerusalem, Jerusalem 91904, Israel}

\begin{abstract}
The position of a reaction front, propagating into a metastable state, fluctuates because of the shot noise of reactions and diffusion. A recent theory [B. Meerson, P.V. Sasorov, and Y. Kaplan,  Phys. Rev. E \textbf{84}, 011147 (2011)] gave a closed analytic expression for the front diffusion coefficient in the weak noise limit. Here we test this theory in stochastic simulations involving reacting and diffusing particles on a one-dimensional lattice. We also investigate a small noise-induced systematic shift of the front velocity compared to the prediction from the spatially continuous deterministic reaction-diffusion equation.
\end{abstract}
\maketitle

\section{Introduction}
\label{intro}

Effects of shot noise on the propagation of reaction fronts have been the subject of extensive studies in physics, chemistry and biology, see e.g. Refs. \cite{vanSaarloos03,Panja}  for reviews. The shot noise comes from the discreteness of reacting particles and random character of their reactions and migration in space. The role of shot noise in the front dynamics can be very different for fronts propagating into an unstable state and for those propagating into a metastable state (a state which is linearly stable but non-linearly unstable).  A front propagating into an unstable state can be extremely sensitive to the shot noise in the leading edge of the front \cite{Panja,Tsimring,Derrida1,vanSaarloos03,Derrida06,MSfisher}. In particular, for pulled fronts, the front diffusion coefficient $D_f$, describing random wandering of the front position, induced by the noise, is only \emph{logarithmically} small with the typical number of particles $N\gg 1$ in the front region \cite{Derrida06}. Fronts propagating into a metastable state behave more ``normally" in this respect, and the scaling $D_f \sim 1/N$  was conjectured a while ago \cite{vanSaarloos03,Panja}. Not much was known beyond this conjecture until recently, when a systematic theory of noisy reaction fronts propagating into a metastable state was developed, and closed analytic expression for $D_f$ derived  \cite{MSK}. In this work we will test the theory of Ref. \cite{MSK} in stochastic simulations
of the microscopic model of reacting and diffusing particles on a one-dimensional lattice. In addition to the front wandering we will also address a more subtle effect of a systematic shift in the \emph{average} front position,
compared to prediction from the reaction-diffusion equation:  a spatially continuum mean-field theory for this system \cite{Murray,Mikhailov}. This systematic shift was previously addressed in Ref. \cite{Khain}, where it was suggested
that it is exponentially small with respect to the carrying capacity of the system $K$, defined below. We revisit
this issue here.

\section{Theory}
\label{theory}

\subsection{General}
\label{general}

The microscopic model we will adopt here, see e.g. Refs. \cite{MSK,Khain}, involves a single species of particles $A$ residing on a one-dimensional lattice of sites. The number of particles $n_i$ on each site $i$ varies in time as a result of two types of Markov processes \cite{Gardiner}. The first type involves  stochastic on-site reactions, such as $A\to 2A$, $A\to 0$, etc. with given rates constants. Assuming for simplicity that, in each of these reactions, the number of particles changes by one, one can reduce all on-site reactions to a birth-death Markov process with given birth and death rates $\lambda(n_i)$ and $\mu(n_i)$, respectively.  The second process is independent random walk of each particle between neighboring sites with rate constant $D_0$.

A spatially continuous deterministic theory for this system can be derived under two assumptions \cite{MSK}. First, assuming that the carrying capacity is a large parameter, $K\gg 1$, one can write
\begin{equation}\label{rates}
    \lambda(n_i)=\nu K \bar{\lambda}(q_i)\;\;\;\mbox{and} \;\;\;\mu(n_i)=\nu K \bar{\mu}(q_i),
\end{equation}
where $n_i\gg 1$, $q_i=n_i/K$, $\bar{\lambda}(q_i)\sim \bar{\mu}(q_i) \sim 1$, and $\nu$ is a rate constant \cite{Doering,AM2010}. When neglecting noise, this brings about a spatially-discrete version of the deterministic reaction-diffusion equation \cite{Keener}
\begin{equation}\label{rateq2}
   \dot{q}_i=\nu f(q_i)+ D_0 (q_{i-1}+q_{i+1}-2 q_i)\,,
\end{equation}
where $f(q_i)=\bar{\lambda}(q_i) - \bar{\mu}(q_i)$ is the rescaled birth-death rate function. Assuming, in addition, a sufficiently fast hopping, $D_0\gg \nu$, one can use a continuous position coordinate $x$ instead of the discrete index $i$ and transform Eq.~(\ref{rateq2}) to the reaction-diffusion equation \cite{Murray,Mikhailov}
\begin{equation}\label{rateeq3}
   \partial_t q=\nu f(q)+D \partial_x^2 q\,,
\end{equation}
where $D=D_0 h^2$ is the diffusion coefficient of the particles, and $h$ is the lattice constant.

A spatially continuous deterministic reaction front corresponds to a traveling wave solution (TWS) of Eq.~(\ref{rateeq3}), $q(x,t)=q_0(\xi)$, $\xi=x-c_0 t$, that connects two linearly stable homogeneous states described by zeros of the function
$f(q)$. Let one of them,  $q=q_*>0$, be located at $x\to -\infty$, and
the other, $q=0$, at $x\to \infty$. There is also a linearly unstable state $q_u$, so that $0<q_u<q_*$.  The TWS $q_0(\xi)$ obeys the equation
\begin{equation}\label{4}
D q_0^{\prime\prime}+c_0 q_0^{\prime}+\nu f(q_0)=0
\end{equation}
and is unique up to an arbitrary shift in $\xi$ \cite{Murray,Mikhailov}. Let us introduce the effective potential $V(q)=\int_0^q f(u) \,du$. Then the state $q=0$ is retreating ($c_0>0$) when $V(q_*)>0$  \cite{Murray,Mikhailov}.  In this case $q=0$ is called the metastable state. For $V(q_*)<0$ the state $q=q_*$ is
retreating ($c_0<0$), and so it is metastable. Finally, for $V(q_*)=0$ the deterministic front is standing, $c_0=0$.  We will call the corresponding set of parameters the stall point. The typical front width is on the order of the diffusion length $l_D=(D/\nu)^{1/2}$.

To account for shot noise, one can go back to the stochastic model on the lattice and use the master equation describing the evolution of the multivariate probability distribution
$P(\mathbf{n},t)= P(n_1,n_2,\dots,t)$. For typical, small fluctuations this master equation can be approximated,  via a truncated expansion,
by a Fokker-Planck equation \cite{MSK,MS}. Instead of dealing with the Fokker-Planck equation directly, Meerson \textit{et al.} \cite{MSK} derived a Langevin equation to which the Fokker-Planck equation is equivalent. For $D_0\gg \nu$, this Langevin equation is continuous in space and takes the form \cite{MSK}
\begin{eqnarray}
\partial_t q(x,t) &=& \nu f(q)+ D \partial_x^2 q+\sqrt{\frac{\nu g(q)\,h}{K}}\, \eta(x,t) \nonumber \\
&+&\partial_x \left[\sqrt{\frac{2 q(x,t)\, Dh}{K}}\, \chi(x,t)\right],
\label{rw040}
\end{eqnarray}
where $g(q)=\bar{\lambda}(q)+\bar{\mu}(q)$ is rescaled \emph{on-site} diffusion coefficient in the space of population sizes, whereas $\eta(x,t)$ and $\chi(x,t)$ are independent Gaussian noises which have zero means and are
delta-correlated in $x$ and in $t$:
\begin{equation}
\left\langle \eta(x,t)\eta(x^{\prime},t^{\prime})\right\rangle=\left\langle \chi(x,t)\chi(x^{\prime},t^{\prime})\right\rangle=\delta(x-x^{\prime})\, \delta(t-t^{\prime}).
\label{rw060}
\end{equation}
The (multiplicative) noise terms in Eq.~(\ref{rw040}) come
from the on-site birth and death processes, and from the random walk (the third and fourth terms on the right, respectively).

Rescaling the position coordinate $x$ and time $t$ by $l_D$ and $1/\nu$, respectively, one can rewrite Eq.~(\ref{rw040}) as
\begin{equation}\label{2}
  \partial_{t} q(x,t) = f(q) +  \partial_{x}^2 q(x,t) + N^{-1/2} R(x,t,q) \,,
\end{equation}
where
\begin{equation}\label{R}
   R(x,t,q)= \sqrt{g(q)}\, \eta(x,t)+\partial_{x}[\sqrt{2q}\, \chi(x,t)]\,.
\end{equation}
As the parameter  $\varepsilon \equiv N^{-1/2} \ll 1$ is small, one can solve Eq.~(\ref{2}) perturbatively  around the deterministic TWS with an a priori unknown front position, slowly varying in time. The first order of this perturbation theory yields a closed-form analytic result \cite{MSK} for the front diffusion coefficient $D_f$ in the frame moving with the average front velocity $\bar{c}$. In the dimensional variables
\begin{equation}\label{Dflin}
    D_f=\frac{D}{s_0N}\,,
\end{equation}
where
\begin{equation}
\label{s}
s_0= \frac{\left[\int_{-\infty}^{\infty} d\xi (q_0^{\prime})^2 e^{c_0 \xi}\right]^2}{\int_{-\infty}^{\infty} d\xi\,\Big\{ \frac{1}{2}\,(q_0^{\prime}e^{c_0\xi})^2\,  g(q_0)+q_0 \left[\left(q_0^{\prime} e^{c_0 \xi}\right)^{\prime}\right]^2\Big\}}
  \,.
\end{equation}
It follows from Eq.~(\ref{Dflin}) and the definition of $N$ that $D_f$ scales as $K^{-1}$ with the carrying capacity parameter $K$, and as $D^{1/2}$
with the particle diffusion coefficient $D$.

The probability distribution of the typical position $X$ of the fluctuating front at time $\tau$  is a Gaussian:
\begin{equation}\label{gauss}
    P(X,\tau)\simeq \left(\frac{s_0N}{4 \pi D \tau}\right)^{1/2} \exp \left[-\frac{s_0 N (X-\bar{c} \tau)^2}{4 D \tau}\right].
\end{equation}
The Gaussian asymptote holds in the region of $|X/\tau-\bar{c}|\ll (\nu D)^{1/2}$ \cite{MSK}. Predictions (\ref{Dflin})-(\ref{gauss}) were also obtained from a WKB approximation to the master equation, in conjunction with linearization of the WKB equations around the deterministic TWS solution \cite{MSK}.

In the first order of the perturbation theory in $\varepsilon$, that the authors of Ref. \cite{MSK} confined themselves to, the average front velocity does not change compared with the prediction of the spatially continuous deterministic theory: $\bar{c}=c_0$. In the second order in $\varepsilon=N^{-1/2}\ll 1$ a small systematic shift $\delta c$ of the average front velocity should appear, caused by the shot noise: $\bar{c}=c_0+\delta c$. In the rescaled variables
$\delta c$ is expected to behave as $\delta c=\chi\,\varepsilon^2$, where $\chi$ is a dimensionless factor, which magnitude and sign are determined by the properties of functions $f(q)$ and $g(q)$, see Eqs. (\ref{2}) and (\ref{R}). In dimensional variables
\begin{equation}\label{shift1}
    \delta c=\chi l_D \nu \varepsilon^2=\frac{\chi \nu h}{K}.
\end{equation}
That is, the shot-noise-induced velocity shift is expected to scale as $K^{-1}$. This is in contrast to the scenario of Ref. \cite{Khain} which predicts  a front velocity shift \emph{exponentially} small in $K$.  Somewhat surprisingly, the front velocity shift (\ref{shift1}) is independent of $D$. The actual second-order perturbative calculations are very cumbersome, and we will not embark on this route here, confining ourselves to the simple argument leading to Eq.~(\ref{shift1}) with the undetermined dimensionless factor $\chi$.

What is the expected validity range of Eqs. (\ref{Dflin}) and (\ref{shift1})? Equation~(\ref{Dflin}) only requires $D_0\gg \nu$ (fast hopping) and $K\gg 1$ (weak
shot noise). Equation~(\ref{shift1}) demands an additional criterion. This is because there is an additional, well-known mechanism of shift of the average front velocity.
It is purely deterministic and comes from the discreteness of the lattice, see Eq.~(\ref{rateq2}) \cite{Keener}. At small  $\nu/D_0$ this velocity shift behaves as $(\delta c)_d= \mbox{const} \times (\nu/D_0) c_0$, as shown by Keener \cite{Keener}. This deterministic velocity shift will mask the shot-noise-induced velocity shift unless $|\delta c| \gg |(\delta c)_d|$.  This criterion can be rewritten as the following strong inequality:
\begin{equation}\label{morecrit}
\frac{D_0 h}{K c_0}\gg 1.
\end{equation}
Far from the stall point the front velocity
$c_0$ can be roughly estimated as $c_0\sim \nu l_D$, and the two criteria, $K\gg 1$ and
Eq.~(\ref{morecrit}) become a strong double inequality
\begin{equation}\label{double}
    1 \ll K \ll  (D_0/\nu)^{1/2},
\end{equation}
that is satisfied for sufficiently fast hopping, $D_0/\nu\gg 1$.

Close to the stall point  $c_0$ is small, and the criterion (\ref{morecrit}) is easily satisfied, making the right inequality in Eq. (\ref{double}) unnecessary (one still needs to demand $D_0\gg \nu$). This feature makes standing fronts, $c_0=0$, especially convenient
for numerical studies of small front velocity fluctuations and shift induced
by shot noise.  Another (even more apparent) advantage of the case of $c_0=0$ is the possibility to
perform the simulations in relatively short systems.

\subsection{Example}
\label{example}

A simple set of on-site reactions which exhibits bistability is  $A \to 0$ and $2A \rightleftarrows 3A$.
Let the rate constants of these three processes be $\mu_0$, $\lambda_0$ and $\sigma_0$, respectively. Following Ref. \cite{MSK}, we will perform rescaling so that
$\bar{\lambda}(q)=2q^2$ and $\bar{\mu}(q)=\gamma\, q+q^3$, where $\nu=3\lambda_0^2/(8\sigma_0)$, $K=3\lambda_0/(2\sigma_0)$, and $\gamma=8\mu_0\sigma_0/(3\lambda_0^2)$.
The on-site dynamics exhibits bistability at $\delta^2 \equiv 1-\gamma>0$. In this case the zeros of function $f(q)=-q\,(q-q_u) (q-q_*)$
describe two stable fixed points of the deterministic theory, $0$ and $q_*=1+\delta$, and an unstable fixed point $q_u=1-\delta$ such that $0<q_u<q_*$. Here the TWS $q_0(\xi)$ is elementary \cite{Murray,Mikhailov}:
\begin{equation}\label{Q0}
    q_0(\xi)=\frac{\delta +1}{1+e^{(1+\delta)\,\xi/(\sqrt{2}\,l_D)}},\;\;\;\;
    c_0= (3 \delta-1)\sqrt{\frac{\nu D}{2}},
\end{equation}
According to the spatially continuous deterministic theory, the state $q=q_*$ is advancing at $1/3<\delta<1$, retreating at $0<\delta<1/3$ and stationary at $\delta=1/3$. The function $g(q)$, entering Eq.~(\ref{s}), is equal to $g(q)=\bar{\lambda}(q)+\bar{\mu}(q)=\gamma\, q+2q^2+q^3$.

In this example, the factor $s_0=s_0(\delta)$ from Eq.~(\ref{s}) can be evaluated analytically for any $\delta$ \cite{MSK}.  In its turn, the factor $\chi$ from Eq.~(\ref{shift1}) also depends only on $\delta$, but this dependence is unknown at present.  In a similar bistable system, studied in Ref. \cite{Khain}, the front velocity shift $\delta c$ even changed sign when a parameter similar to our $\delta$ was varied. In the present work we focus on the particular value of $\delta=1/3$, corresponding to the stall point of the mean field theory. Here $s_0=\sqrt{2}/6$ \cite{MSK},
$\lambda_0=9\mu_0/(2K)$, $\sigma_0=27\mu_0/(4K^2)$ and $\nu=9\mu_0/8$, and we can rewrite Eqs.~(\ref{Q0}), (\ref{Dflin}), (\ref{gauss}) and (\ref{shift1}) as
\begin{eqnarray}
   q_0(\xi)&=&\frac{4}{3\left(1+e^{\sqrt{\frac{\mu_0}{D}}\xi}\right)},\;\;\;\;\xi=x-\delta c \, t,\label{Q00}\\
   D_f&=& \frac{9 h \sqrt{\mu_0 D}}{2K},\label{Df0}\\
   P(X,\tau)&=& \frac{1}{\sqrt{4\pi D_f \tau}}\,\exp\left[-\frac{(X-\delta c \,\tau)^2}{4D_f \tau}\right],\label{gauss0}\\
   \delta c&=&\alpha\,\frac{\mu_0 h}{K}, \label{shift2}
\end{eqnarray}
where Eq.~(\ref{shift2}) includes an unknown numerical coefficient $\alpha={\cal O}(1)$.
It is Eqs.~(\ref{Q00})-(\ref{shift2}) that we aimed at verifying in stochastic simulations.

\section{Simulations}

We performed extensive Monte Carlo simulations on a one-dimensional lattice with spacing $h=1$ and integer length $L\gg 1$. Every lattice site $0\leq i < L$ can be occupied by any number of particles $n_i$. The number of particles changes, at each step,
because of independent random walk or one of the reactions $A \to 0$ and $2A \rightleftarrows 3A$,
see Sec. \ref{example}. We employed the Gillespie algorithm \cite{Gillespie}. A site $j$ is chosen with probability proportional to $n_j$. Then a particle on that site is chosen at random. It can jump to a neighboring site on the right or left, produce a new particle, or die with probabilities
\begin{eqnarray}
p_{right} &=& p_{left} = D/(b_1 +d_1+2D), \\ \nonumber
p_{birth} &=& b_1/(b_1 +d_1+2D), \\ \nonumber
p_{death} &=& d_1/(b_1 +d_1+2D),
\label{ratesMC}
\end{eqnarray}
respectively. Here
$$
b_1 = \lambda_0 (n_j-1)/2 \,\,\, \mbox{and} \,\,\, d_1 = \mu_0 + \sigma_0(n_j-1)(n_j-2)/6.
$$
After every single-particle event, the time is advanced by
$1/[M(b_1 +d_1+2D)]$, where $M=\sum_j n_j$ is the total number of particles in the system. At the boundaries $j=0$ and $L$, particles are not allowed to migrate out of the system, which corresponds to no-flux boundary conditions. We took the initial number density profile that corresponds to the mean-field front solution given by Eq. (\ref{Q00}): $n_j(t=0) = K\,q_0(j)$, such that the front interface is located in the center of the system, $X_0=L/2$.

\label{simulations}
\begin{figure}[ht]
\includegraphics[width=2.5 in,clip=]{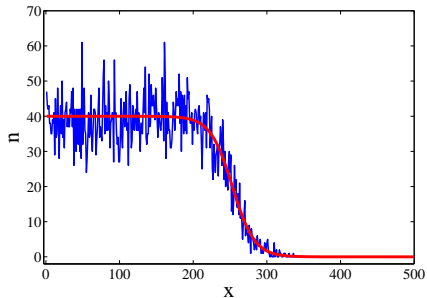}
\caption{(Color online). A typical realization of the front density profile observed in a Monte Carlo simulation of the lattice model with reactions $A \to 0$ and
$2A \rightleftarrows 3A$ and unbiased random walk. The smooth line shows the mean-field solution $n(j) = K\,q_0(j)$, see Eq.~(\ref{Q00}), with a fitted front position. The parameters are: $K=30$, $D=20$, $\nu = 0.1$, $\delta = 1/3$, and $h=1$.}
\label{profile}
\end{figure}

\begin{figure}[ht]
\includegraphics[width=2.5 in,clip=]{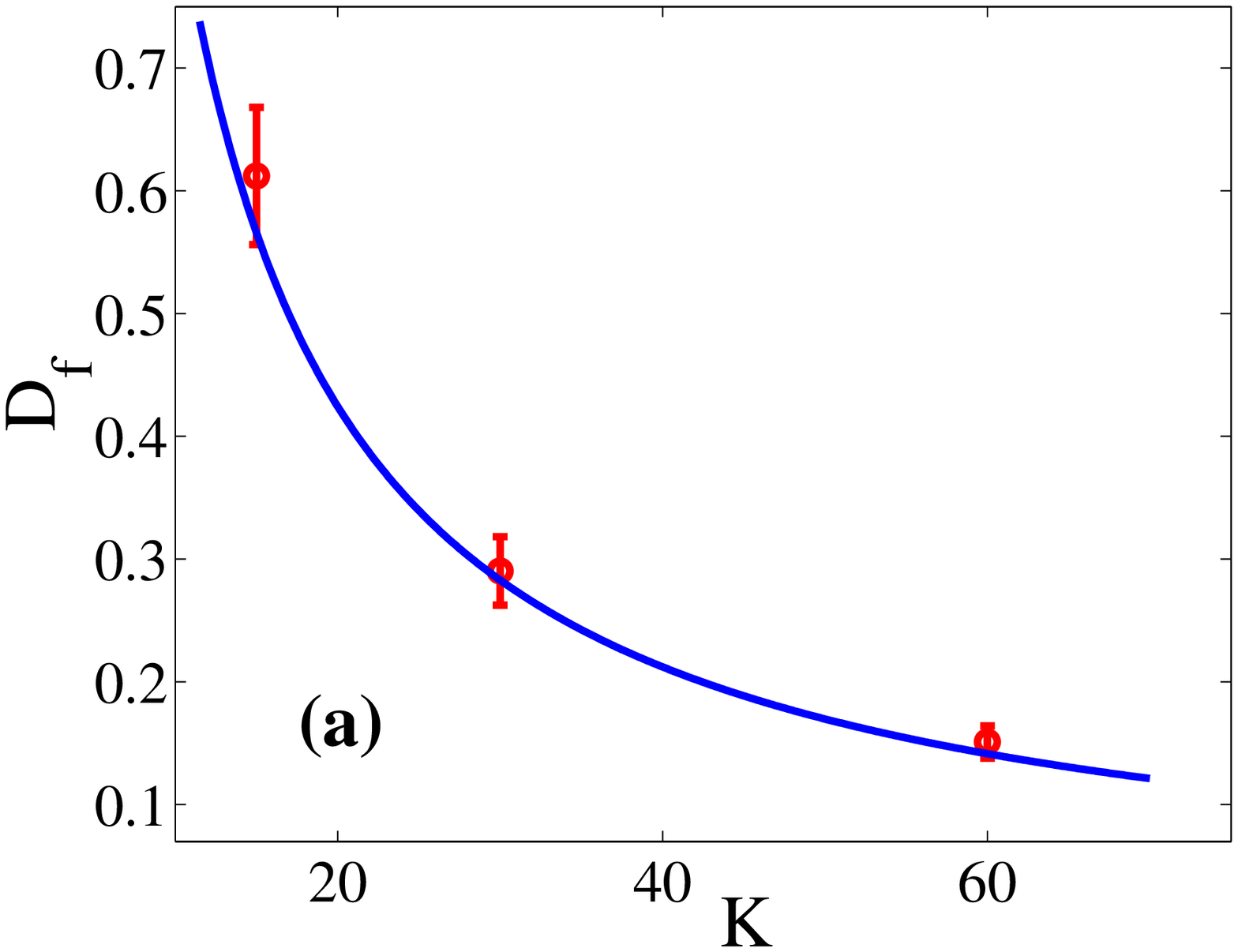}
\includegraphics[width=2.5 in,clip=]{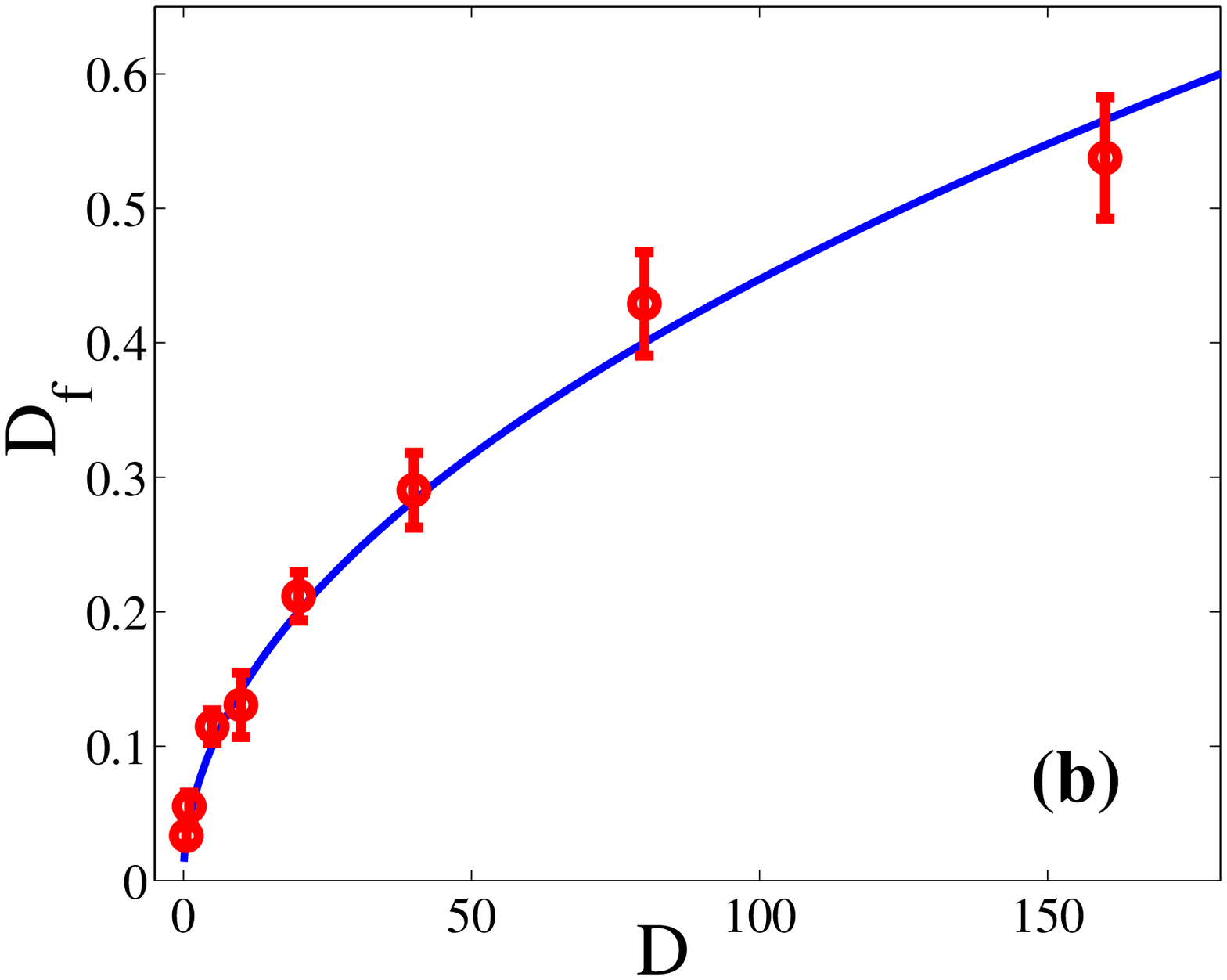}
\caption{(Color online). The front diffusion coefficient $D_f$ as a function of $K$ (a) and $D$ (b). Symbols with error bars: the simulation results. The solid lines are computed from Eq.~(\ref{Df0}). The parameters are: $D=40$ (a) and $K=30$ (panel b). Also, $\nu = 0.1$, $\delta = 1/3$, the simulation time is $\tau = 2000$, and the system size is $L=600$.}
\label{Df}
\end{figure}

We performed many simulations
and determined the front position for different values of parameters. The number of simulations for each value of $K$ and $D$ varied between $60$ and $420$. A typical density profile $n_j$, observed in the simulations, is shown in Fig. \ref{profile}. To find the front position $X$, we fitted the density $n_j$ by the mean-field profile that moved distance $X-X_0$; the value of $X$ served as the fitting parameter. For a fixed simulation time $\tau$, stochastic fronts corresponding to different realizations propagate a different distance $X-X_0$. We computed the standard deviation $\sigma_{st}$ of the distribution of the front positions $X$, and determined the fronts diffusion coefficient
$D_f=\sigma^2_{st}/(2\tau)$. Figures \ref{Df} a and b show the simulation results for $D_f$ as a function of the carrying capacity $K$ and of the diffusion coefficient of the particles $D$, respectively. These plots also show theoretical predictions from Eq.~(\ref{Df0}), without any fitting parameters. The agreement is good in both cases.

\begin{figure}[ht]
\includegraphics[width=2.5 in,clip=]{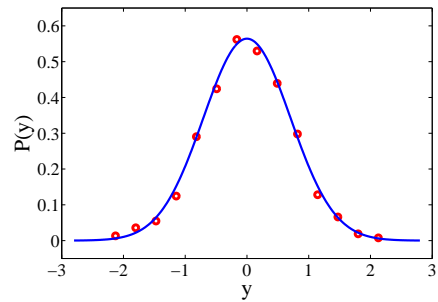}
\caption{(Color online). Symbols: the rescaled distribution of front positions measured in simulations for different values of $D$ and $K$ (the total of $1591$ simulations).
The solid line:  the normal distribution $\bar{P}(y) = \pi^{-1/2}\exp(-y^2)$ predicted by theory \cite{MSK} for typical fluctuations. No fitting parameters are used. The rest of parameters are: $\nu = 0.1$, $\delta = 1/3$, the simulation time is $\tau = 2000$, the system size is $L=600$.}
\label{dist}
\end{figure}

Furthermore, we combined our sets of measurements for different values of parameters to obtain a rescaled probability distribution of the front positions, and therefore of the average front velocities.  First, in each set of simulations with the same parameters, we transformed to a moving frame, $X^{\prime}=X- \bar{X}$, by shifting the measured distribution by the average displacement $\bar{X}$ of the fronts in this set of simulations. Then we combined the data from the different sets of simulations on a single plot that shows the rescaled probability distribution of $X^{\prime}$, $\bar{P}(y) = \sqrt{4 D_f \tau} P(X^{\prime})$, versus the rescaled front position
$$
y = X^{\prime} \, \sqrt{\frac{s_0 N}{4Dt}} = \frac{X^{\prime}}{\sqrt{4D_f\tau}},
$$
see Eq.~(\ref{gauss0}). This plot is shown in Fig. \ref{dist}. As one can see, there is a very good
collapse of data for different sets of simulations and a good agreement with the theory.

\begin{figure}[ht]
\includegraphics[width=2.5 in,clip=]{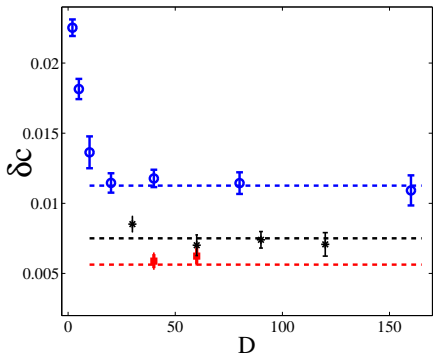}
\caption{(Color online). The noise-induced velocity shift $\delta c=\bar{X}/\tau$ as a function of $D$ for three values of the carrying capacity: $K=60$ (circles), $K=90$ (asterisks) and $K=120$ (squares). The rest of parameters are the same as in Fig. \ref{dist}.}
\label{deltac}
\end{figure}

\begin{figure}[ht]
\includegraphics[width=2.5 in,clip=]{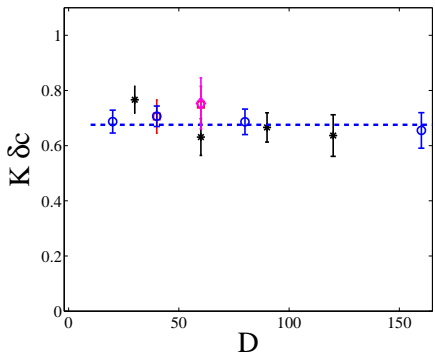}
\caption{(Color online). The rescaled noise-induced velocity shift $K \delta c$ as a function of $D$ for four values of the carrying capacity: $K=60$ (circles), $90$ (asterisks), $120$ (squares) and $150$ (a single point, denoted by diamond). The rest of parameters are the same as in Fig. \ref{dist}. The dashed line corresponds to $K \delta c = 0.68$. Then, using Eq.~(\ref{shift2}) with $\mu_0=(8/9) \nu$, we obtain
$\alpha\simeq 7.6$.}
\label{rescaled}
\end{figure}

We have also measured the average shift in the front position $\bar{X}$, with the aim of verifying  the scaling relation (\ref{shift2}) for $\delta c$. To remind the reader, for $\delta=1/3$, the front velocity, predicted by the mean field theory, is zero. The measured average shift, therefore, comes from the shot noise. Figure \ref{deltac} shows the front velocity correction $\delta c=\bar{X}/\tau$ as a function of the particle diffusion coefficient $D$ for three different carrying capacities $K$. One can see that, at sufficiently large $D$, $\delta c$ becomes independent of $D$, as predicted by  Eq.~(\ref{shift1}). This equation also predicts that, in the large-$D$ limit, the velocity correction must decrease with $K$ as $1/K$. This prediction is fully supported by our measurements, see Fig. \ref{rescaled}, where the rescaled quantity $K \delta c$ is plotted versus $D$, and good collapse of data points for $K=60, 90$ and $120$, and a single data point for $K=150$, is observed. For the case of $\delta=1/3$, considered here, the coefficient $\alpha$ in Eq.~(\ref{shift2}) is about $7.6$.

\section{Summary}
Our stochastic simulations strongly support theoretical prediction of Ref. \cite{MSK} of the diffusion coefficient $D_f$
of noisy reaction fronts propagating into a metastable state.
The inverse dependence of $D_f$ on $K$ and the $\sqrt{D}$ dependence of $D_f$ on $D$ are both verified, as well
as the Gaussian distribution of the (typical) front positions in the reference frame moving with the average front speed.

The simulations also clearly show that, for sufficiently high hopping rates, the noise-induced systematic
shift of the front velocity is a second order effect in the small parameter $\varepsilon\equiv 1/\sqrt{N}$, and so it scales as $K^{-1}$.

The front diffusion coefficient comes from the Gaussian region of the fluctuations of the front position. These are typical, small fluctuations.  The WKB theory of Ref. \cite{MSK} also dealt with rare large fluctuations  and predicted
non-Gaussian tails of the front velocity distribution. Observing
these tails in stochastic simulations requires a more specialized simulation algorithm and is left for future work.

\subsection*{Acknowledgments}
We thank David A. Kessler, Leonard M. Sander and Pavel V. Sasorov for useful discussions. B.M. was supported by the Israel Science Foundation (Grant No. 408/08) and by the U.S.-Israel Binational Science Foundation
(Grant No. 2008075).


\begin{thebibliography}{99}
\bibitem{vanSaarloos03} W. van Saarloos, Phys. Rep. \textbf{386}, 29 (2003).
\bibitem{Panja} D. Panja, Phys. Rep. \textbf{393}, 87 (2004).
\bibitem{Tsimring} L.S. Tsimring, H. Levine, and D.A. Kessler,
Phys. Rev. Lett. \textbf{76}, 4440 (1996).
\bibitem{Derrida1} \'{E}. Brunet, B. Derrida,  Phys. Rev. E
\textbf{56}, 2597 (1997).
\bibitem{Derrida06} \'{E}. Brunet and B. Derrida, Comput. Phys. Commun. \textbf{121}–\textbf{122},
376 (1999); J. Stat. Phys. \textbf{103}, 269 (2001); \'{E} Brunet, B. Derrida,
A. H. Mueller, and S. Munier, Phys. Rev. E \textbf{73}, 056126 (2006).
\bibitem{MSfisher} B. Meerson and P.V. Sasorov, Phys. Rev. E \textbf{84}, 030101 (R) (2011).
\bibitem{MSK} B. Meerson, P.V. Sasorov, and Y. Kaplan,  Phys. Rev. E \textbf{84}, 011147 (2011).
\bibitem{Murray} J. D. Murray, \textit{Mathematical Biology. I: An Introduction} (Springer, New York, 2003).
\bibitem{Mikhailov} A.S. Mikhailov, \textit{Foundations of Synergetics I. Distributed Active Systems} (Springer-Verlag, Berlin, 1990).

\bibitem{Khain} E. Khain, Y. T. Lin, and L. M. Sander, Europhys. Letters \textbf{93}, 28001 (2011).
\bibitem{Gardiner} C.W. Gardiner, \textit{Handbook of Stochastic Methods}
(Springer Verlag, Berlin, 2004).
\bibitem{Doering} C.R. Doering, K.V. Sargsyan, and L.M. Sander, Multiscale Model. and Simul. \textbf{3}, 283 (2005).
\bibitem{AM2010} M. Assaf and B. Meerson, Phys. Rev. E \textbf{81}, 021116 (2010).
\bibitem{Keener} J. P. Keener, SIAM J. Appl. Math. \textbf{47}, 556 (1987).

\bibitem{MS} B. Meerson and P. V. Sasorov, Phys. Rev. E \textbf{83}, 011129 (2011).

\bibitem{Gillespie} D.T. Gillespie, J. Phys. Chem. \textbf{81},  2340 (1977).


\end{thebibliography}
\end{document}